\documentstyle[12pt,epsfig]{article}
\def\bge{\begin{equation}}
\def\ene{\end{equation}}
\def\bg{\begin{eqnarray}}
\def\en{\end{eqnarray}}

\def\aleq{\stackrel{<}{\sim}}

\begin{document}
\renewcommand{\thefootnote}{\fnsymbol{footnote}}
\begin{flushright}
ADP-97-7/T245 
\end{flushright}
%
%\vspace{0.5cm}
%
\begin{center}
\begin{LARGE}
Rho-meson mass in light nuclei
\end{LARGE}
\end{center}
\vspace{0.5cm}
\begin{center}
\begin{large}
K.~Saito\footnote{ksaito@nucl.phys.tohoku.ac.jp} \\
Physics Division, Tohoku College of Pharmacy \\ Sendai 981, Japan \\
K.~Tsushima\footnote{ktsushim@physics.adelaide.edu.au} and 
A.~W.~Thomas\footnote{athomas@physics.adelaide.edu.au} \\
Department of Physics and Mathematical Physics \\
and \\
Special Research Center for the Subatomic Structure of Matter \\
University of Adelaide, South Australia, 5005, Australia
\end{large}
\end{center}
\vspace{0.5cm}
\begin{abstract}
The quark-meson coupling (QMC) model is applied to a study of 
the mass of the $\rho$-meson in helium and carbon nuclei.
The average mass of a $\rho$-meson formed in $^{3,4}$He and $^{12}$C 
is expected to be around 730, 690 and 720 MeV, respectively.  
\end{abstract}
%Keywords: Relativistic mean-field theory, finite nuclei, hadron masses, 
%quark degrees of freedom, MIT bag model
%
%%%%%%%%%%%%%%%%%%%%%%%%%%%%%%%%%%%%%%%%%%%%%%%%%%%%%%%%%%%%%%%%%%%%%
%
\newpage
% INTRODUCTION

As the nuclear environment changes, hadron properties are nowadays expected to 
be modified~\cite{sai1,brown,hatkun,pr1,sai2,hatsuda}.  
In particular, the variation of the light vector-meson mass 
is receiving a lot of attention, both theoretically and experimentally.  
Recent experiments from the HELIOS-3, CERES and NA50 
collaborations at SPS/CERN energies have shown that there exists a large 
excess of the lepton pairs in central S + Au, S + W, Pb + Au and Pb + Pb 
collisions~\cite{exp}.  An anomalous $J/\psi$ suppression in Pb + Pb 
collisions has also been reported by the NA50 collaboration~\cite{j/psi}. 
Those experimental results may give a hint of some change of hadron 
properties in nuclei (for a recent review, see Ref.~\cite{qm96}).  
We have previously studied the variation of hadron 
masses in medium mass and heavy nuclei using the quark-meson coupling (QMC) 
model~\cite{pr1,pr2}. 

On the other hand, even in light nuclei like helium and carbon, 
an attempt to measure the $\rho^0$-meson mass in the nucleus is underway 
at INS, using tagged photon beams and 
the large-acceptance TAGX spectrometer at the 1.3 GeV Tokyo Electron 
Synchrotron~\cite{ins}.  
They have measured $\rho^0$ decay into two charged pions 
with a branching fraction of approximately 
100\% in low-atomic-number nuclei, in which 
pions suffer less from final state interactions.  The actual experiments 
involved measurements of the $\pi^+ \pi^-$ photoproduction on 
$^3$He, $^4$He and $^{12}$C nuclei in the energy region close to the $\rho^0$ 
production threshold. 
In view of this experimental work it is clearly 
very interesting 
to report on the variation of the $\rho$-meson mass in these light nuclei. 

% THE QUARK-MESON COUPLING MODEL

To calculate the hadron mass in a {\em finite} nucleus, 
we use the second version of the quark-meson coupling (QMC-II) model, 
which we have recently 
developed to treat the variation of hadron properties in nuclei (for 
details, see Ref.~\cite{pr2}). 
This model was also used to calculate detailed properties of spherical, 
closed shell nuclei from $^{16}$O to $^{208}$Pb, where it was shown that 
the model can reproduce fairly well the observed charge density 
distributions, neutron density distributions, etc.~\cite{np1}.  
In this approach, which began with work by Guichon~\cite{guichon} in 
1988, quarks in non-overlapping nucleon bags interact 
{\em self-consistently} with scalar ($\sigma$) and 
vector ($\omega$ and $\rho$) mesons (the latter also being described by 
meson bags),  
in the mean-field approximation (MFA).  
Closely related investigations have been made in 
Refs.~\cite{jen,mil,xxx}.

In the actual calculation, we use the MIT bag model in static, spherical 
cavity approximation. 
The bag constant $B$ and the parameter $z_N$, in the familiar form of the 
MIT bag model Lagrangian~\cite{mit}, are fixed to reproduce the free 
nucleon mass 
($M_N$ = 939 MeV) and its free bag radius ($R_N$ = 0.8 fm). 
Furthermore, to fit the free vector-meson masses, 
$m_{\omega}$ = 783 MeV and $m_{\rho}$ = 770 MeV, we introduce new 
$z$-parameters for them, $z_{\omega}$ and $z_{\rho}$.  
Taking the quark mass in the bag to be  
$m_q$ = 5 MeV, we find $B^{1/4}$ = 170.0 MeV, $z_N$ = 3.295, 
$z_\omega$ = 1.907 and $z_\rho$ = 1.857~\cite{pr2}.  

The model has several coupling constants to be determined: 
the $\sigma$-nucleon 
coupling constant (in free space), 
$g_{\sigma}$, and the $\omega$-nucleon coupling constant, $g_{\omega}$, 
are fixed to fit the binding energy 
($-15.7$ MeV) at the correct saturation density ($\rho_0 = 0.15$ fm$^{-3}$) for 
symmetric nuclear matter.  Furthermore, the $\rho$-nucleon coupling constant, 
$g_\rho$, is used to reproduce the bulk symmetry energy, 35 MeV.  Those 
values are listed in Tables 1 and 3 of Ref.~\cite{pr2}.

Within QMC-II, the non-strange vector mesons 
are described by the bag model and their masses in the 
nuclear medium are 
given as a function of the mean-field value of the $\sigma$ meson at that 
density~\cite{pr2}.  
However, the $\sigma$ meson itself is not so readily represented  
by a simple quark model (like a bag), because it couples strongly 
to the pseudoscalar ($2 \pi$) channel and a direct 
treatment of chiral symmetry in medium is important~\cite{hatkun}.  
On the other hand, many approaches, including 
the Nambu--Jona-Lasinio model~\cite{hatkun,bern}, the Walecka 
model~\cite{sai1,cail} and Brown-Rho scaling~\cite{brown} suggest that the 
$\sigma$-meson mass in medium, $m_{\sigma}^{\star}$, should   
be less than the free 
one, $m_\sigma$. We have parametrized it using a quadratic 
function of the scalar field: 
\bge
\left( \frac{m_{\sigma}^{\star}}{m_{\sigma}} \right) = 1 - a_{\sigma} 
(g_{\sigma} \sigma) + b_{\sigma} (g_{\sigma} \sigma)^2 , 
\label{sigmas}
\ene
with $g_\sigma \sigma$ in MeV.  
To test the sensitivity of our results to the $\sigma$ mass in the medium, 
the parameters were chosen~\cite{pr2}: 
($a_\sigma$ ; $b_\sigma$) = (3.0, 5.0 and 7.5 $\times 10^{-4}$ MeV$^{-1}$ ; 
10, 5 and 10 $\times 10^{-7}$ MeV$^{-2}$) for sets A, B and C, respectively.
These values lead to a reduction of the $\sigma$ mass for sets A, B and C 
by about 2\%, 7\% and 10\% 
respectively,  
at saturation density.  

Using this parametrization for the $\sigma$ mass, 
the $\rho$-meson mass in matter is found to take quite a simple form 
(for $\rho_B \aleq 3 \rho_0$): 
\bge
m_\rho^{\star} \simeq m_\rho - \frac{2}{3} (g_\sigma \sigma) \left[ 1 - 
  \frac{a_\rho}{2} (g_\sigma \sigma) \right] , 
\label{vm}
\ene
where $a_\rho \simeq 8.59, 8.58$ and $8.58 \times 10^{-4}$ (MeV$^{-1}$) 
for parameter sets A, B and C, respectively~\cite{pr2}.  

% NUMERICAL RESULTS

For medium and heavy nuclei, it should be reasonable to use the MFA,  
and the mean-field values of all the meson fields at position ${\vec r}$ 
in a nucleus can be determined by (self-consistently) solving a set of 
coupled non-linear differential equations, generated from the QMC-II 
Lagrangian density~\cite{pr2}.  
We have calculated the $\rho$-meson mass in $^{12}$C in that way.  
However, for $^3$He and $^4$He, the MFA is not expected to be reliable.
Therefore, we shall use a simple local-density approximation to 
calculate $m_\rho^{\star}$ in helium.  

In practice it is easy to parametrize the mean-field value of the $\sigma$ 
field calculated in QMC-II as a function of 
$\rho_B$ (see Fig.1 of Ref.~\cite{pr2}) and it is given as 
\bge
g_{\sigma} \sigma \simeq s_1 x + s_2 x^2 + s_3 x^3, 
\label{sigval}
\ene
where $x = \rho_B / \rho_0$ and the parameters, $s_{1 \sim 3}$, are listed 
in Table~\ref{sval}.  
Therefore, once one knows the density distribution of the helium nucleus, 
one can easily calculate $g_\sigma \sigma$ at position ${\vec r}$ from 
Eq.(\ref{sigval}), 
and then calculate $m_\rho^{\star}(r)$ in the nucleus using Eq.(\ref{vm}). 

In this paper we use a simple gaussian form for the density distribution of 
$^3$He, in which the width parameter, $\beta_3$, is fitted to reproduce the 
rms charge 
radius of $^3$He, 1.88 fm.  For $^4$He, we parametrized 
the matter density as: 
\bge
\rho_4 (r) = A_4 ( 1+ \alpha_4 r^2) \exp (-\beta_4 r^2), 
\label{heli4}
\ene
where $\alpha_4$ = 1.34215 (fm$^{-2}$) and $\beta_4$ = 0.904919 (fm$^{-2}$). 
This was chosen to reproduce the rms matter radius of $^4$He, 1.56 fm, 
and the measured central depression in the charge density. 

\begin{figure}[hbt]
\begin{center}
%\epsfile{file=3he.ps,height=11cm}
\epsfig{file=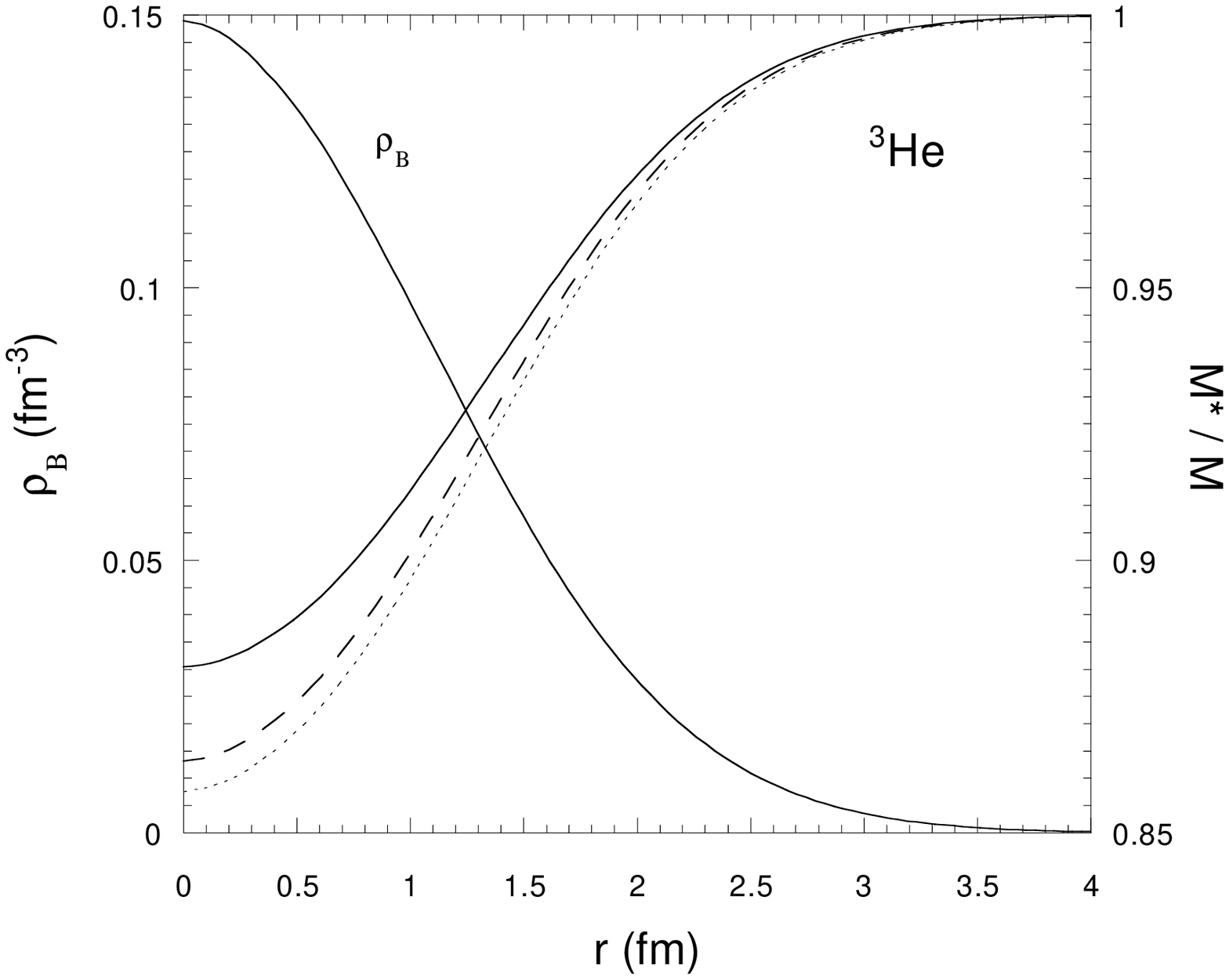,height=11cm}
\caption{}
\label{fig:3he}
\end{center}
\end{figure}

\begin{figure}[hbt]
\begin{center}
%\epsfile{file=4he.ps,height=11cm}
\epsfig{file=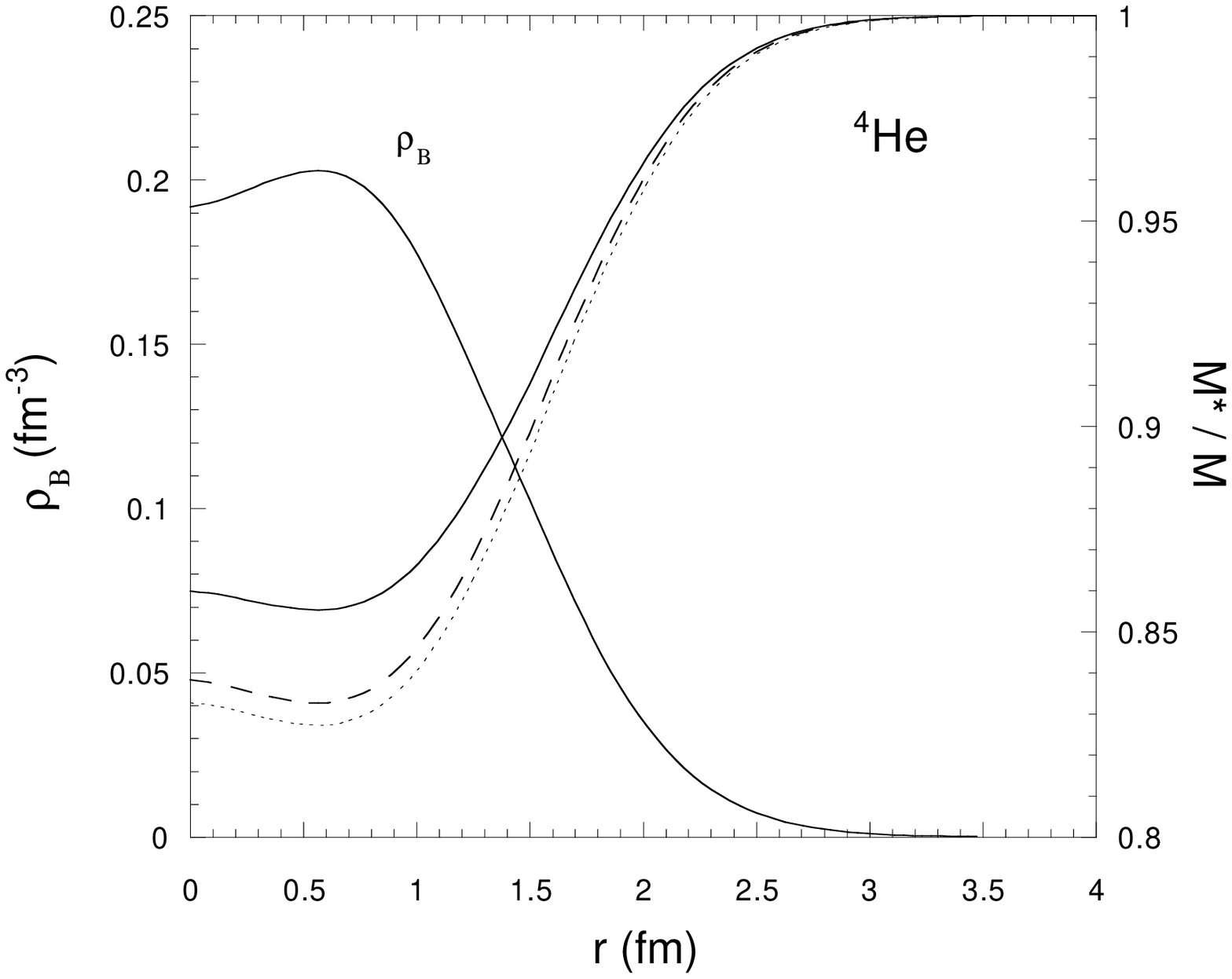,height=11cm}
\caption{}
\label{fig:4he}
\end{center}
\end{figure}

\begin{figure}[hbt]
\begin{center}
%\epsfile{file=12C.ps,height=11cm}
\epsfig{file=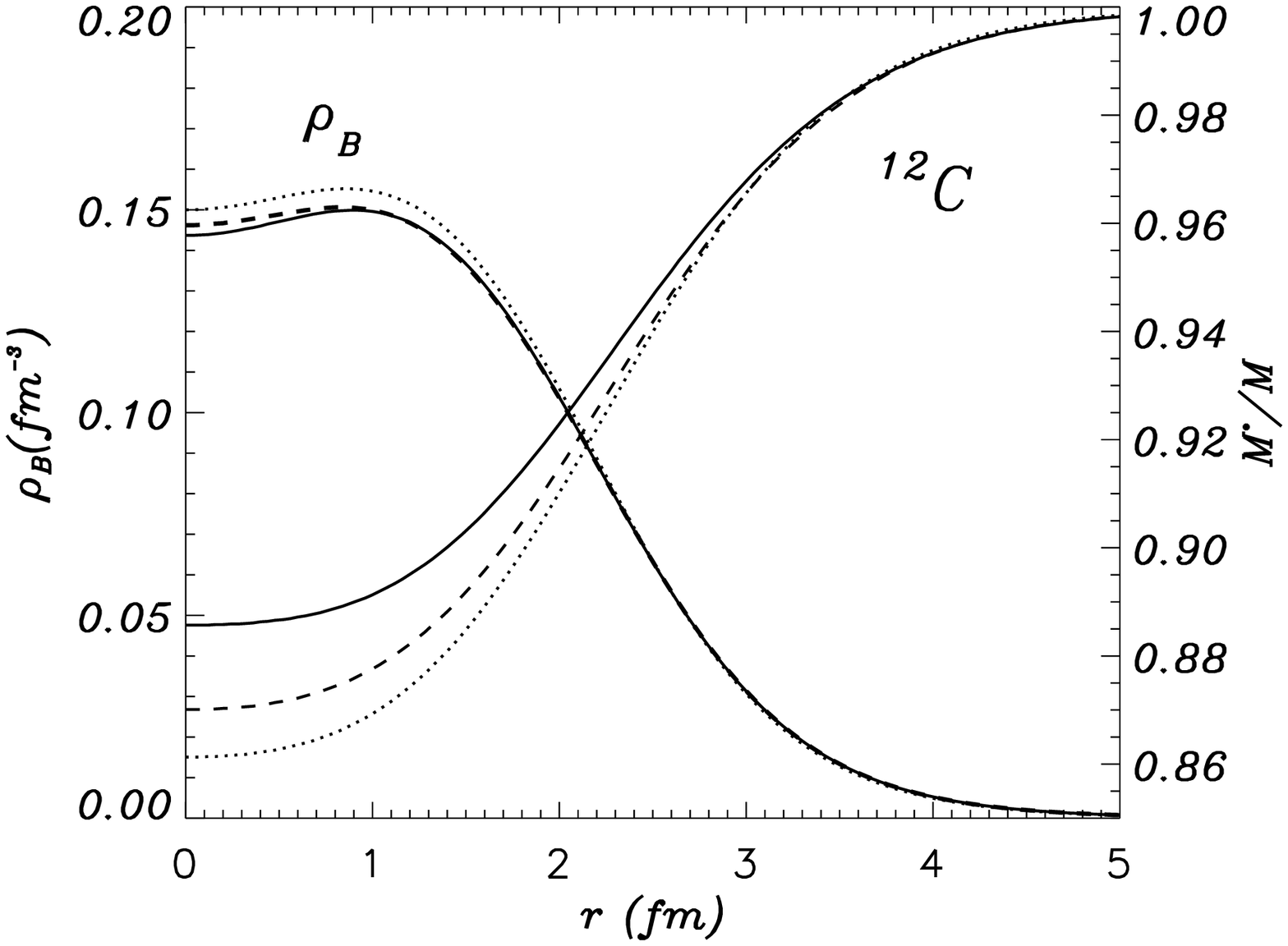,height=11cm}
\caption{}
\label{fig:12c}
\end{center}
\end{figure}

Now we show our numerical results.  In Figs.~\ref{fig:3he}$-$\ref{fig:12c}, 
the density distributions and the $\rho$-meson masses in $^3$He, 
$^4$He and $^{12}$C are illustrated (for $^{3,4}$He 
the density distribution is common to all of the 
parameter sets, A $\sim$ C).  
The $\rho$-meson mass decreases by about 10 $\sim$ 15 \% at the center of 
the nucleus, although it depends a little 
on the parameter set chosen for the $\sigma$ mass variation. 

We also show the average $\rho$-meson mass in the nucleus, which is defined 
as
\bge
\langle m_\rho^{\star} \rangle_A = \frac{1}{A} \int d{\vec r} \ \ 
\rho_A(r) m_\rho^{\star}(r), 
\label{avmas}
\ene
where $\rho_A(r)$ is the density distribution of the nucleus A. 
The average mass is summarized in Table~\ref{avm}. 
In the present model the $\rho$-meson mass seems to be reduced by about 
40 MeV in $^3$He, 80 MeV in $^4$He and 50 MeV in $^{12}$C, due to the nuclear 
medium effect. The larger shift in $^4$He is a consequence of the higher 
central density in this case.

It may be also very interesting to study the variation of the width of the 
$\rho$ meson in a nucleus.  Unfortunately, since the present model does not 
involve the effect of the width, we cannot say anything about it. 
Asakawa and Ko~\cite{asa}, however, have reported on the mass and width of 
the $\rho$ meson although their calculations were carried out in nuclear 
matter.  They have used a realistic spectral function, 
which was evaluated in the vector dominance model including the effect of the 
collisional broadening due to the $\pi$-N-$\Delta$-$\rho$ dynamics, on the 
hadronic side of the QCD sum rules, and concluded that the width of the $\rho$ 
meson decreases {\em slightly} as the density increases, which implies that 
the phase space 
suppression (from the $\rho \to 2 \pi$ process) due to the reduction of the 
$\rho$-meson mass more or less balances the collisional 
broadening at finite density.  Provided that the width of the $\rho$ meson 
is not significantly decreased by such medium corrections 
we may expect that the $\rho$ meson created by an external 
beam should decay inside the nucleus. This should lead to  
a clean signal of the variation of the $\rho$-meson mass~\cite{hatsuda}. 

% CONCLUSION

In conclusion, we have calculated the $\rho$-meson mass in 
$^3$He, $^4$He and $^{12}$C using the QMC-II model, and 
found that it is reduced by about 10 $\sim$ 15 \% in those nuclei. 
It will be very interesting to compare our results with the experimental 
data taken at INS and currently being analysed~\cite{ins}. 

\vspace{2cm}
This work was supported by the Australian Research Council.  The authors 
thank G. Lolos for valuable discussions and comments. One of the 
authors (K.S.) thanks K. Maruyama for stimulating discussions.  
%
%%%%%%%%%%%% Bibliography %%%%%%%%%%%%%%%%%%%%%%%%%%%%%%%%%
%

%
\newpage
\begin{large}
\flushleft{\underline{Figure captions}}
\end{large}
\begin{description}
\item[Fig.1] 
Effective $\rho$-meson mass and the density distribution in 
$^3$He. The solid, dashed and dotted curves are, respectively, for the 
parameter sets A, B and C. 
\item[Fig.2] 
Same as for Fig.1 but for $^4$He. 
\item[Fig.3] 
Same as for Fig.1 but for $^{12}$C. 
\end{description}
\newpage
\newpage
\newpage
\newpage
\begin{table}[htbp]
\begin{center}
\caption{Three parameters for the mean-field value of $\sigma$ (in MeV). }
\label{sval}
\begin{tabular}[t]{c|ccc}
\hline
type & $s_1$ & $s_2$ & $s_3$ \\
\hline
 A &  195.2 & $-$52.1 & 5.1 \\
 B &  214.0 & $-$44.3 & 1.9 \\
 C &  228.0 & $-$51.8 & 2.8 \\
\hline
\end{tabular}
\end{center}
\end{table}
\newpage
\begin{table}[htbp]
\begin{center}
\caption{Average $\rho$-meson mass (in MeV). }
\label{avm}
\begin{tabular}[t]{c|ccc}
\hline
type & $^3$He & $^4$He & $^{12}$C \\
\hline
 A &  732 & 701 & 723 \\
 B &  727 & 691 & 718 \\
 C &  725 & 688 & 715 \\
\hline
\end{tabular}
\end{center}
\end{table}

\end{document}